\documentstyle[twocolumn,aps,epsfig]{revtex}
\begin{document}
\draft
\title{Charged Scalar Fields in an External \\
Magnetic Field: Renormalisation and Universal Diamagnetism}
\author{ Debnarayan Jana \footnote{E-mail: deb@rri.ernet.in}}
\address{Raman Research Institute, Bangalore 560 080, India}
\date{\today}
\maketitle
\begin{abstract}
The physical and mathematical mechanism behind diamagnetism of  N (finite) 
spinless
bosons (relativistic or non-relativistic) is well known. The
mathematical
signature of this diamagnetism follows from Kato's inequality while 
its physical way of
understanding goes back to Van Leewen.
 One can guess that it might be true in the field theoretic case also.
  While the work on
systems with a finite number of degrees of freedom suggests that the same
result is true in a field theory, it does not by any means prove it.  In
the field theoretic context one has to develop a suitable regularisation
scheme to renormalise the free energy. We show  that charged scalar fields
in (2+1) and (3+1) dimensions  are always diamagnetic, even in the presence of 
interactions and at finite
temperatures. This generalises earlier work on the diamagnetism of charged
spinless bosons to the case of infinite degrees of freedom. We also
discuss possible applications of the theory.
\end{abstract}
\section{Introduction} 
A classical gas of charged point particles is non-magnetic, by Van
Leeuwen's Theorem~\cite{van}. But quantum mechanically, free spinless Bose
particles in a uniform magnetic field show diamagnetism\cite{la1}. One
finds that  the energy of the system in a magnetic field is higher than
in the absence of the field. Simon\cite{sim} proved quite generally that
the ground state  of  non-relativistic spinless bosons interacting through
an arbitrary potential {\it always increases} in a  magnetic field. He
went on to extend this result  by showing that the free energy in the
presence of a magnetic field is always greater than the free energy in the
absence of a magnetic field at all temperatures~\cite{sim}. An alternative
proof of this result is given in~\cite{ss}. All this  work  which deals
with systems with a finite number of degrees of freedom suggests that
diamagnetism is a universal property of spinless bosons. In field theory
(which describes systems with an infinite number of degrees of freedom)
charged spinless bosons are described by complex scalar fields. One might
therefore expect that charged scalar fields would also show diamagnetic
behaviour. In fact Brydges et. al.~\cite{dav} have shown this diamagnetic
inequality for the interacting scalar fields on a lattice. This result
insisted on an ultraviolet cutoff. Here we want to extend to the continuum
case with proper account of the renormalization. With this motivation we 
study the magnetic behavior of scalar
field theories in both (2+1) and (3+1) dimensions \footnote{ At this point 
we would like to emphasize 
that it is extremely naive and dangerous to conclude the response 
of the field theory from the corresponding N particle
systems. For example, N fermions at $T=0$ show Pauli 
paramagnetism while its counter
field theory QED vacuum is  diamagnetic.}.\\

The paper is organised as follows. The first part deals with finite
temperature free  scalar field theory in the presence of an external
homogeneous magnetic field. Here, we explicitly calculate the partition
function and the free energy as a function of the applied 
magnetic field in (2+1) dimension.
This expression is formally divergent. Using a suitable regularization
scheme we compute the {\it difference} in the free energy (with and
without the magnetic field) and obtain a finite answer. This difference is
also shown to be positive, thus establishing the diamagnetic behaviour of
free charged scalar fields. We establish the renormalisation of the free
energy through numerical and analytical calculations. Next we discuss the
renormalization of the free energy in (3+1) dimension. In this connection
we also show the quantitative difference between the responses in odd and
even spatial dimension. More specifically, we point out the drastic change
of the behaviour of the free energy at strong magnetic field  in the 
zero temperature limit.\\

We then move on to interacting scalar field theory in the second part.
Here, we cannot evaluate the partition function explicitly. Nevertheless
we prove the universal diamagnetism of scalar fields by assuming a finite
momentum cutoff in the theory. If the theory is renormalizable, then one
can take this cutoff to infinity while maintaining finiteness of all
physical quantities. In both  cases the results obtained are exact. In fact
we present a {\it non-perturbative} approach to prove  this diamagnetic 
inequality for this interacting case in an arbitrary spatial dimension. 
 Before
the conclusion, we illustrate few examples of spinless bose systems and
point out the variation of the susceptibility with temperature above the
critical temperature in a physical way.\\

Although this work is rederivation of a known result, we hope that it has
a value beyond the purely pedagogical. Here in most of the paper we will talk
about the response of the vacuum.
 The vacuum as a medium has been a fruitful
picture in trying to develop intuition about abelian and non-abelian gauge
theories. In fact we point out several interesting points regarding the 
nature of the vacuum say for example QED and QCD in presence 
of an external magnetic
field.\\

\section{Free case}
\subsection{ (2+1) dimension}
\par
In this subsection we calculate the free energy of free scalar fields in the
presence of an external uniform magnetic field. For ease of presentation 
we work here first in two spatial dimensions. The interesting physics
takes place in the plane  normal to the applied  field. In the next subsection
we will discuss the renormalization of free energy in (3+1) dimension.

Let $\Phi$ be a complex scalar field which describes charged spinless Bosons.
 The Lagrangian density of a free charged scalar field in the
 presence of a constant homogeneous external magnetic field is given by
 
  \begin{equation}
 {\cal L}=(D_{\mu}\Phi)^\ast(D^\mu\Phi) - m^2(\Phi^\ast\Phi)
 \end{equation}
    where $\mu=0,1,2$,
 \begin{equation}
  D_{\mu} = \partial_{\mu}-{\rm i} e A_{\mu}                          
 \end{equation}
 and $m$ and $e$ are the mass and charge respectively.
 (We set $\hbar=1$ and $c = 1$).
  Now, we write the complex field in term of two real fields $\Phi_{1}$
 and $\Phi_{2}$.
 \begin{equation}
 \Phi=\frac{\Phi_{1}+i\Phi_{2}}{\sqrt 2}\nonumber,
 \Phi^\ast=\frac{\Phi_{1}-i\Phi_{2}}{\sqrt 2}
 \end{equation}
 This theory has a global U(1) symmetry and therefore a conserved Noether 
 charge ${\rm Q}$, 
 given by
 \begin{equation}
  Q=\int d\,^2x~ (\Pi_{1}\Phi_{2}-\Phi_{1}\Pi_{2})
  \end{equation}
   where
   \begin{equation}
    \Pi_{i}=\partial_{0}\Phi_{i}
   \end{equation} 
 
 The Hamiltonian density of the system is given by
 \begin{eqnarray}
  {\cal H} &=&\frac{1}{2}(\Pi_{1}^2+\Pi_{2}^2)+\frac{1}{2}(\nabla\Phi_{1})^2+
  \frac{1}{2}(\nabla\Phi_{2})^2+\nonumber\\
           & &~~~~~~\frac{1}{2}(m^2+e^2A^2)(\Phi_{1}^2+\Phi_{2}^2)-
\vec{j}\cdot \vec{A},
 \end{eqnarray}
 where the current density j is given by
 \begin{equation}
 \vec{j}= {\rm -e}~(\Phi_{1}\vec{\nabla}\Phi_{2}-\Phi_{2}\vec{\nabla}\Phi_{1}).
 \end{equation}
 We now suppose that the external magnetic field is uniform in the $x-y$ plane.
 
We choose the temporal gauge ( $A_{0}=0$).
   The constant magnetic field $B$ is 
     \begin{equation}
    B = \partial_{x}~A_{y}- \partial_{y}~A_{x}
   \end{equation}
 where $A_x$ and $A_y$ are independent of t. 
 
  The action of this theory is 
 
\begin{equation} 
 S =\int_{0}^{\beta} \int d^{2}x~ d\tau ~ {\cal L}(\Phi,\Phi^\ast,A),
 \end{equation}
where  $ \tau$ is the imaginary time variable which runs 
  from 0 to $\beta$ (=1/($k_{B}T$)), the inverse temperature. 
 The action defined above is quadratic and so the partition function can
be evaluated exactly.
 As is usual in finite temperature field theory~\cite{ftft}, we impose
 periodic boundary conditions for  Bosonic fields
 \begin{equation}
      \Phi({\bf x},0) = \Phi({\bf x},\beta).
 \end{equation}

 Now, the partition function of this theory can be written as
  \begin{eqnarray}
 Z(B)&=&\int {\cal D}[\Pi_{1}] {\cal D}[\Pi_{2}]\int {\cal D}[\Phi_{1}]
     {\cal D}[\Phi_{2}] \exp\left[ \int d\tau~d^2x \right. \nonumber \\
     & &\left.\left({\rm i}\Pi_{1}\frac{\partial\Phi_{1}}{\partial\tau}+  {\rm
    i}\Pi_{2}\frac{\partial\Phi_{2}}{\partial\tau}
    -H(\Phi_{1}, \Phi_{2}, \Pi_{1}, \Pi_{2})\right.\right.\nonumber \\
     & &~~~~\left.\left.+\mu(\Phi_{2}\Pi_{1}-\Phi_{1}\Pi_{2})\right)\right]
\end{eqnarray}

 Here $\mu$ is the chemical potential associated with the conserved
charge $Q$. The  charge density ($Q/A$) has to be contrasted with the 
usual number density.
The charge density refers here to the difference between the particle density
and the antiparticle density and hence can take any sign while the number
density by definition is always positive. In that sense $\mu$ is not the usual
chemical potential used in Grand Canonical Ensemble. We pick the gauge in which the vector potential {\bf A} is (-By,
0), and expand the complex scalar field in terms of modes adapted to the
present situation.  These modes solve the  Klein-Gordon equation in an
external magnetic field. The eigenfunctions are labelled by one discrete
($l$) and one continuous $p_{x}$ quantum number and the spectrum depends
on $l$ only.  In the gauge we choose, the modes are plane waves  in the x
direction and harmonic oscillator (i.e. gaussian) wavefunctions in the y
direction.
 
The spectrum is given by
 \begin{equation}
  \omega_{l}^2 =  m^2 + (2l+1)eB,~~~ l=0,1,2.......\infty
  \end{equation}
The degeneracy of these states for fixed $l$ is ${e A B}/{2\pi}$ ,
where $A$ is the area of the system.  So, these modes can be thought of
as quantized harmonic oscillators. Expanding
the fields $\Phi_{1}$ and $\Phi_{2}$ in these modes the system reduces to
a collection of harmonic oscillators with frequency $\omega_{l}$.

By standard manipulations ~\cite{ftft}, we get the free energy as

\begin{eqnarray}
F(B)&=&{\rm 2\pi eA}B\sum_{l=0}^\infty\left[\omega_{l}+
     \frac{1}{\beta}\ln(1-{\rm exp}(-\beta(\omega_{l}-\mu)) \right.\nonumber\\
    & &\left.+\frac{1}{\beta} \ln(1-{\rm exp}(-\beta(\omega_{l}+\mu))\right]
 \label{sq1} 
 \end{eqnarray}

The first term in the square brackets corresponds to the zero point
fluctuation  of the vacuum and the other two terms are  finite
temperature contributions of the particles and antiparticles respectively.
Without the above formalism one can also calculate the vacuum energy by 
summing up the zero point energies for the modes of all the fields. Of 
course when we sum over all modes the energy actually diverges. One has
to introduce a regularisation scheme or a cutoff so that the vacuum energy
becomes finite.\\

It is easy to notice that this zero point energy is divergent due
to the summation of infinity number of modes (Landau levels). In
conventional field theory this infinite zero point energy is always
discarded; since it can be reabsorbed in a suitable redefinition of the
zero of energy. This is justified in the sense that the infinite zero point
energy is unobservable. However, the change in zero point energy caused by
the external constraint is finite and observable. So, according to Casimir's
~\cite{cas} idea, the physical vacuum energy can be defined as the difference
between the zero point energy corresponding to the vacuum configurations 
with constraints and the one corresponding to the free vacuum configurations.
This definitions must be supplemented in general with a regularisation
prescription in order to obtain a finite final convergent expression.
We will first compare the free energy of the system with and without the 
magnetic field at zero temperature.\\

\subsubsection{Zero Temperature Case} 
The free energy of the system in presence of the magnetic field at zero
temperature is given by
\begin{equation}
F_{0}(B)=2\pi A~eB\sum_{l=0}^\infty~\omega_{l},\label{cx}
\end{equation}
where $\omega_{l}^2=m^2+(2l+1)eB$. Obviously, this sum diverges. In order
to obtain a finite answer we need to impose a cutoff $L$ in the sum
(\ref{cx}). Then the free energy becomes
\begin{equation}
F_{0}(B,L)=2\pi A~eB\sum_{l=0}^L~\omega_{l}.\label{tre}
\end{equation}
The free energy in the absence of the magnetic field at zero temperature
is  given by the divergent expression
\begin{equation}
F_{0}(0)= 2\pi A\int_{0}^\infty~p dp~\sqrt{(p_{x}^2+p_{y}^2)+m^2}\label{vx}
\end{equation}
We regularize this expression by imposing a cutoff $\Lambda$. Then the free
energy (\ref{vx}) becomes
\begin{equation}
F_{0}(0,\Lambda)= 2\pi
A\int_{0}^\Lambda~p dp~\sqrt{(p_{x}^2+p_{y}^2)+m^2}\label{frt}
\end{equation}
In order to compare the free energies in equations (\ref{tre}) and
(\ref{frt}), we choose the cutoffs $L$ and $\Lambda$ in such a way that
both systems have the same number of modes. Such a procedure can be
justified on physical grounds if one imagines that  the magnetic field
is turned on adiabatically~\cite{com}. In the Appendix A and B, we 
show that this
mode matching scheme is the correct regularisation procedure. More 
specifically, in Appendix A we compare this scheme with another scheme
which looks apparently correct and in Appendix B we justify this mode
matching scheme from the field theoretic point of view. 

Counting the modes upto the L-th Landau level we find
\begin{equation}
2\pi A~eB\sum_{l=0}^L~~{\rm 1} = 2\pi A~eB(L+1)
\end{equation}
Similarly, for the momentum cutoff upto $\Lambda$ we get the modes without
the magnetic field as
\begin{equation}
2\pi A\int_{0}^\Lambda pdp = \pi~A\Lambda^2\label{op}
\end{equation}
Equating these gives us
\begin{equation}
\Lambda^2=2~eB~(L+1)\label{mode1}
\end{equation}
Now, the free energy in absence of the magnetic field depends on magnetic
field through the momentum cutoff and is given by
\begin{equation}
F_{0}(0,L)=2\pi A\int_{0}^{\Lambda(B)}~pdp~\sqrt{p^2+m^2}
\end{equation}
The difference between the two free energies is given by
\begin{equation}
\Delta F(B,L)=F_{0}(B,L)-F_{0}(0,L)
\end{equation}
We define
$f(B)={F_{0}(B,L)}/{2\pi A}$, $\tilde{f}(B)={F_{0}(0,L)}/{2\pi A}$ and
$\Delta f(B)=f(B)-\tilde{f}(B)$. Numerically evaluating these sums one
can show that for finite $L$, $\Delta f(B)$ the difference 
between two large quantities is positive.
As the cutoff $L$ goes to infinity,  $\Delta f(B)$ becomes the
difference between two infinities. In this limit we find that $\Delta
f(B)$ tends to a finite value. Thus, the susceptibility at zero
temperature in the relativistic case is non-zero. This vacuum
susceptibility can be interpreted as due to virtual currents.

We now show analytically that $\Delta F(B)$ is positive i.e. the vacuum is
diamagnetic.  Note that
\begin{equation} 
\Delta f(B)~=~f(B)-\tilde{f}(B)=\sum_{l=0}^\infty~a_{l}(B,m) \label{ok}
\end{equation} 
where $a_{l}(B,m)$ is given by
\begin{eqnarray}
a_{l}(B,m)&=&eB\left[\sqrt{m^2+(2l+1)eB}-\right.\nonumber\\
& &\left.\int_{0}^{1}~d\alpha~\sqrt{m^2+2(l+\alpha)eB}\right]
\end{eqnarray}
Introducing a dimensionless quantity $\rho=\frac{eB}{m^2}$ the above
equation becomes
\begin{equation}
a_{l}(\rho)=\rho
\left[\sqrt{1+(2l+1)\rho}-\int_{0}^{1}~d\alpha~\sqrt{1+2(l+\alpha)\rho}\right]
\end{equation}

The positivity of $a_{l}(\rho)$ for each $l$ can be proved geometrically.
Defining $z_{l}={(1+2l\rho)}/{2\rho}$ and $f(\alpha)=\sqrt{z_{l}+\alpha}$, the
coefficient  $a_{l}(\rho)$ can be rewritten in terms of $c_{l}(\rho)$ as
\begin{equation}
c_{l}(\rho)=
\frac{a_{l}(\rho)}{\sqrt{2}~\rho^{3/2}}=f(1/2)-\int_{0}^1~d\alpha~f(\alpha).
\end{equation}
Since, the function $f(\alpha)$ is convex, the area under the tangent
drawn at $\alpha=1/2$ is greater than the area under the curve (see the 
figure below).
\begin{figure}
\hspace{2cm}
\epsfig{file=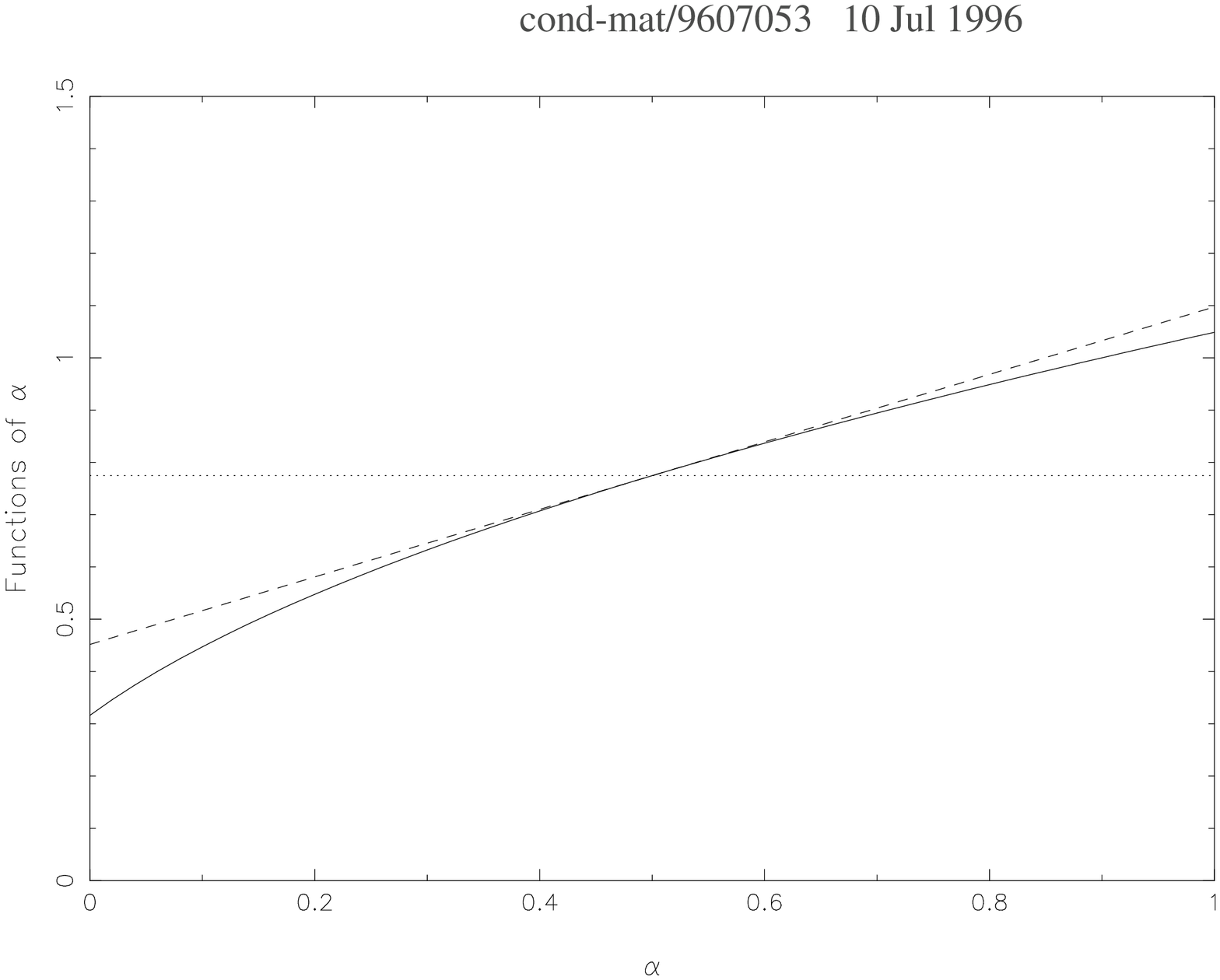,width=10cm,height=8cm,angle=-90}
Figure 1: The full line is the curve $f(\alpha)=\sqrt{.1+\alpha}$. The
dashed line is the tangent to the above curve at $\alpha=.5$. It is
 straightforward to see that the area under the curve is less than the
area under the tangent. Also the area under the tangent is the same as the
area of the rectangle. The area of this rectangle is given by $f(1/2)\times 1
= f(1/2)$. Hence, the positivity of $a(.1)$ is proved. This can be 
generalised to any positive value of $z_l$.

\label{fig:plot}
\end{figure}

This shows that $c_{l}(\rho)$ is positive. To show the
convergence of the sum (\ref{ok}) we note that
\begin{equation}
\int_{0}^1~d\alpha~f(\alpha)~\leq~\left[f(1/2)~-~\frac{f(0)+f(1)}{2}\right]
\end{equation}
Now, applying mean value theorem twice one can easily show that 

\begin{equation}
\int_{0}^1~d\alpha~f(\alpha)\leq~\frac{1}{16(z_{l}+\alpha)}{3/2}
\end{equation}
Thus the coefficient $c_{l}(\rho)$ is positive for each
$l$ and the sum converges, hence the diamagnetic inequality is established.
\subsubsection{Massless limit}
In this section we want to discuss the behaviour of the leading term of
the free energy in the massless limit and its consequences. It is easy to
see from the zero temperature free energy that the magnetisation  in the
zero mass limit is given by
\begin{equation}
{\rm M(B)}\sim~-\sqrt{B}
\end{equation}
So, the susceptibility in this zero mass limit is given by
\begin{equation}
\chi(B)~\sim~-~\frac{1}{\sqrt{B}}
\end{equation}
which diverges~\cite{div} as $B$ goes to zero. This divergence of the
susceptibility is reminiscent of the fact that the free energy $F(B)\sim
B^{3/2}$. This variation of the free energy in the  massless limit can
also be understood from dimensional arguments as follows. Since we are
working in natural units $\hbar=1$ and $c=1$, then $[m]\sim[L]^{-1}$.
Then the free energy density (i.e. per unit area ) varies as $[L]^{-3}$.
However, the dimension of $B$ is $[L]^{-2}$. So, the massless limit
restricts the free energy density variation with magnetic field $B$ to
$B^{3/2}$ only. This feature of the susceptibility has already been noticed
in the magnetised pair Bose gas~\cite{mag}. This divergence of the
susceptibility can be compared with that of ideal charged bose gas at the
condensation point~\cite{sca}. The magnetic susceptibility is defined by
\begin{equation}
M=\chi B
\end{equation}
where $B$ is the applied field; what is actually computed, however, is a
quantity $\chi'$ defined by
\begin{equation}
M=\chi' B'
\end{equation}
where $B'$ is the ``acting" field. The relation between $B$ and $B'$
requires special attention; in fact if the ``acting" field is identified
with the average microscopic field then we have for this case
\begin{equation}
B'=B+2\pi M
\end{equation}
This equation immediately gives a relation between $\chi$ and $\chi'$ as
\begin{equation}
\chi=\frac{\chi'}{1-2\pi\chi'}
\end{equation}
This shows that on approaching $B\rightarrow 0$ limit, where
$\chi'\rightarrow -\infty$, $\chi\rightarrow -\frac{1}{2\pi}$ so that the
permeability tends to zero.
So, in this case the external
field will be totally expelled. This happens because of large number of
virtual particle and antiparticle produced in the ground state so that the
overall diamagnetism of the system is high enough to totally expel the
external field. That the magnetisation is non-analytic in the limit
$B\rightarrow 0$ is the actual signature of a phase transition as a
function of external field strength.  Note that this behaviour is
equivalent to the behaviour of the free energy in strong magnetic field
limit. This can be understood  the way $m$ and $B$
appears in $\omega=\frac{eB}{m}$ which is the only energy scale in the
problem. Hence, one should expect the same behaviour as $m\rightarrow 0$
or $B\rightarrow \infty$ because in both situations higher Landau levels
become less and less occupied. In fact in this situation one can construct
the theory in the lowest Landau level.\\

\par
 Another interesting point  is that
there exists a critical magnetic field below which magnetic field will be
totally expelled. This critical field can be estimated as follows. The
effective magnetic field can be defined as $B_{eff}=~B + 2\pi~M$. Here, the
magnetisation $M$ varies as $-~const\sqrt{B}$.  Therefore, there exists a
critical magnetic field $B_{c}$ where the effective field $B_{eff}$
vanishes.
\subsubsection{Finite Temperature Case}
\par
Now, for the finite temperature case  one can regulate the
free energy through the same mode matching regularisation method. Finally,
 one can write down the free energy difference in dimensionless form as
before 
\begin{equation} 
\Delta F(B)= F(B)-F(0)=\sum_{l=0}^\infty~b_{l}(\rho,\delta,\zeta) 
\end{equation} 
where,
\begin{equation}
b_{l}(\rho,\delta,\zeta)=
\frac{\rho}{\delta}
\left[g(\rho,l,1/2)-\int_{0}^{1}~d\alpha~g(\rho,l,\alpha)\right].
\end{equation}
The dimensionless variables are defined as
$\delta=\beta m$ and $\zeta=\beta\mu$.

The coefficient $g(\rho,l,\alpha)$ is given by
\begin{eqnarray}
g(\rho,l,\alpha)&=&
\log\left(1-\exp(-\delta(\sqrt{1+2(l+\alpha)\rho}-\zeta))\right)+\nonumber\\
& &\log\left(1-\exp(-\delta(\sqrt{1+2(l+\alpha)\rho}+\zeta))\right)\label{gb}.
\end{eqnarray}
Now, defining $z_{l}=\frac{\delta^2(1+2l\rho)}{2\rho}$ we can write the
equation (\ref{gb}) 
\begin{eqnarray}
 g(\rho,l,\alpha)&=&
\log\left(1-\exp(-(\sqrt{z_{l}+\alpha}-\zeta))\right)+\nonumber\\ &
&\log\left(1-\exp(-(\sqrt{z_{l}+\alpha}+\zeta))\right).
\end{eqnarray}
The function $g(\rho,l,\alpha)$ is convex,  so the zero temperature
argument applies unchanged. It follows that
the free energy satisfies the following inequality
\begin{equation}
F(B) \geq F(0)
 \end{equation} 
 Thus the  response of the system to the magnetic
  field will be diamagnetic.

\subsection{(3+1) dimension}
 
In this subsection we would like to renormalize the free energy in
presence of an external uniform magnetic field in (3+1) dimension. For
the sake of repetition we do not present here the derivation of the free
energy in (3+1) dimension which can be easily be done in the framework
presented earlier for (2+1) dimension case. In fact  the final result can be
understood easily from the dispersion relation (energy spectrum) for (3+1)
case which is given by
\begin{equation}
\omega_l=\sqrt{p_z^2+m^2+(2l+1)eB},~~~l=0,....\infty
\end{equation}

Thus, we get the free energy of charged
scalar fields in 3d in presence of a constant homogeneous external magnetic
field as

\begin{eqnarray}
F(B)&=&{\rm 2\pi eV}B\sum_{l=0}^\infty\int_{-\infty}^{\infty} dp_z 
\left[\omega_{l}+
     \frac{1}{\beta}\ln(1-e^{-\beta(\omega_{l}-\mu)}) \right.\nonumber\\
    & &\left.+\frac{1}{\beta} \ln(1-e^{-\beta(\omega_{l}+\mu)})\right]
 \label{sq} 
 \end{eqnarray}

From the above equation (\ref{sq}) we notice that the zero point free energy 
is given by
\begin{equation}
F(B,m)=2\pi eBV \sum_{l=0}^{\infty}\int_{-\infty}^{\infty} dp_z
\left[\sqrt{m^2+p_z^2+(2l+1)eB}\right]\label{dh}
\end{equation}
 We regularise this
free energy  by the prescription given in~\cite{la}
which was used in case of QED in an external magnetic field. In principle,
 one can use  the prescription through a novel mode 
matching already developed for (2+1) dimension to regularise this free energy
. But we use this regularisation scheme to compute quantitative
evaluation of the regularised free energy in two interesting limits in the
zero temperature case. As it is clear from (2+1) dimension case that
we cannot evalute the response function quantatively that's why we choose this
scheme for that reason.\\

In field theory there are various regularisation schemes for example
Schwinger's proper time regularisation scheme~\cite{jul} or invariant
regularisation scheme due to Pauli and Villars~\cite{pauli}. Through this
simple regularisation scheme we demonstrate the renormalisation of field
strength and charge as shown in the context of vacuum
polarisation~\cite{jul}.\\

\subsubsection{Zero Temperature Case} 
The divergence of the above integral (Eq. (\ref{dh})) is eliminated 
by subtracting the
value of the sum when $B=0$. Hence, to carry out this renormalisation
procedure it is appropriate to calculate first the convergent expression.
Defining the free energy density $f=F/V$ we can write
\begin{equation}
\frac{\partial f}{\partial(m^2)}=2\pi eB\int_{0}^{\infty}\sum_{l=0}^{\infty}
\left[\frac{dp_z}{\sqrt{m^2+p_z^2+(2l+1)eB}}\right] 
\end{equation}

Differentiating once again we get $\Theta=\frac{\partial^2
f}{\partial (m^2)^{2}}$
\begin{equation}
\Theta=-\pi eB\int_{0}^\infty
\left[\sum_{l=0}^{\infty}
(m^2+p_z^2+(2l+1)eB)^{-3/2}\right] dp_z
\end{equation}
Notice that the integral is now absolutely convergent. Performing the integral
and the sum, we obtain

\begin{eqnarray}
\Theta &=&-\pi
eB\left[\sum_{l=0}^{\infty}\frac{1}{m^2+(2l+1)eB}\right]\nonumber\\ &=&\pi
eB\int_{0}^{\infty} e^{-(m^2+eB)\eta}\left[\frac{1}{1-e^{-2eB\eta}}\right]
d\eta,
\end{eqnarray}
which reduces to a simple integral form given by
\begin{eqnarray}
\Theta&=&\frac{\pi eB}{2}\int_{0}^{\infty}~ \frac{d\eta~~ e^{-m^2\eta}}{\sinh
(eB\eta)}\nonumber\\
    &=&\frac{\pi eB}{2}~\int_{0}^{\infty}~ d\eta~ e^{-m^2\eta}~~ g(eB\eta),
\end{eqnarray}
where $g(eB\eta)=1/\sinh(eB\eta)$.
Now to get back the free energy we have to integrate $\Theta$ twice.
Integrating once we get with a undetermined constant 
\begin{eqnarray}
\Theta_1 &=&\int~\Theta~d(m^2)\nonumber\\
       &=&\frac{\pi eB}{2}~\int_{0}^{\infty}(-1/\eta)~ e^{-m^2\eta}~ g(eB\eta)
~d\eta + C_2.
\end{eqnarray}
Similarly integrating once more one gets back 
\begin{eqnarray}
\Theta_2 &=&\int~ \Theta_1~d(m^2)\nonumber\\
       &=&\frac{\pi eB}{2}~\int_{0}^{\infty}~\frac{1}{\eta^2} e^{-m^2\eta}
g(eB\eta)~~ d\eta \nonumber\\
& &~~~~~~ + C_2~m^2 + C_1
\end{eqnarray}
These two undetermined constants depend only on $B$ but not on $m^2$. Now
it is easy to take $B\rightarrow 0$ limit in the above  formula. This gives
\begin{equation}
\Theta_2
(B=0)=\frac{\pi}{2}~\int_{0}^{\infty}~\frac{1}{\eta^3}~e^{-m^2\eta} d\eta
\end{equation}
Here we have assumed that two constants $C_1$ and $C_2$ vanish as
$B\rightarrow 0$. Therefore the difference between the two free energies
can be written in an integral form as
\begin{eqnarray}
\Delta f
&=&\frac{\pi}{2}~~\int_{0}^{\infty}~~\frac{1}{\eta^3}~~e^{-m^2\eta}~
d\eta\left[\frac{eB\eta}{\sinh(eB\eta)}-1\right]\nonumber\\ &
&~~~~~~~~~~~~~~+C_2~ m^2 +C_1
\end{eqnarray}         
Till now the constants $C_1$ and $C_2$ are undetermined. These are
determined from the following considerations.
Parity and dimensional analysis give us a particular form \cite{scale}
of $\Delta f$
given by
\begin{equation}
\Delta f= m^4~g\left(\frac{B^2}{m^4}\right)\label{dim}
\end{equation}

From equation (\ref{dim}) it is clear immediately that there will be no
odd terms of $m^2$ in the expression for $\Delta f$. Hence, $C_2$ is zero.
Now to find out the coefficient $C_1$ we follow the following prescription.
Notice that the expansion of $\Delta f$ in $B^2$ begins with a term linear
in $B^4$.  Therefore, a term in $B^2$ would alter the coefficient in the
Lagrangian $L_0=\frac{B^2}{8\pi}$. This essentially modifies the
definition of charge. Hence, the elimination of $B^2$ term thus
corresponds to a renormalization of charge.\\

Expanding $\sinh(eB\eta)$ in leading order upto $B^2$ we get
\begin{equation}
C_1=\frac{\pi}{2}\int_{0}^{\infty}
\frac{ e^2B^2\eta^2~e^{-m^2\eta}~d\eta}{6\eta^3}
\end{equation}
Rescaling $m^2\eta\rightarrow \eta$ and defining $b=\frac{eB}{m^2}$
finally we obtain
\begin{equation}
\Delta f=\frac{\pi m^4}{2}\int_{0}^{\infty} e^{-\eta}\frac{d\eta}{\eta^3}
\left[\frac{b\eta}{\sinh(b\eta)}-1+\frac{1}{6}~b^2\eta^2\right]\label{main}
\end{equation}
{\it This renormalized energy difference $\Delta f$ is one of the main
results of the paper}.
Notice that the difference of free energy $\Delta f$ goes to zero as
$B\rightarrow 0$ as it should be since we have subtracted the zero field
free energy.
One can easily show through numerical integration that this free energy is
positive definite and it converges. Before we end this subsection we want to
make a remark on this regularisation scheme. A careful reader will immediately
notice that this scheme will not work in (2+1) dimension. This shows the 
important role played by the $p_z$ integral. 
 In the next two sections we consider two
interesting limits of the free energy difference.\\
\subsubsection{Small Magnetic Field Limit}
 At small magnetic field ($b\eta$ being very small) we can approximate the
free energy difference upto the leading factor in $b^4$ as
\begin{eqnarray}
\Delta f&=&\frac{7\pi m^4 b^4}{720}\int_{0}^{\infty}~e^{-\eta}~\eta~d\eta\nonumber\\
~~~~~~~~&=&\frac{7\pi e^4B^4}{720m^4}
\end{eqnarray}
Thus the free energy difference at small magnetic field limit goes as
$B^4$. This is obvious from the fact that we have subtracted the leading
divergence of $B^2$ term from the free energy. So, the next order term in
the free energy difference is $B^4$ according to symmetry. Notice that the
free energy difference is positive definite and thus establishes the
diamagnetism of the charged scalar fields at zero temperature.
\subsubsection{Strong Magnetic Field Limit}
In this section we would like to address various responses of the system
in the strong magnetic field limit.\\

In a strong magnetic
field ($b\gg 1$)  rescaling $b\eta\rightarrow
\eta$, we get the free energy difference  from equation (\ref{main})
\begin{equation}
\Delta f=\frac{\pi~m^4~b^2}{2}\int_{0}^{\infty}~
e^{-\eta/b}~\frac{d\eta}{\eta^3}
\left[\frac{\eta}{\sinh(\eta)}-1+\frac{1}{6}\eta^2\right]
\end{equation}
When $b\gg 1$, the important range in this integral is $1\ll\eta\ll b$, in
which $e^{-\eta/b} \approx 1$, then terminating the range of integration
(with logarithmic accuracy) at $\eta\approx 1$ and $\eta\approx b$, we get
\begin{equation}
\Delta f\approx \frac{\pi m^4 b^2}{12}\ln b\label{str}
\end{equation}
It is important to note that in the massless limit the free energy in 3d turns
out~\cite{qed} as
\begin{equation}
\Delta f\approx \frac{\pi m^4 b^2}{12}\ln\left(\frac{\overline{\Lambda}^2}{eB}\right)
\end{equation}
where $\overline{\Lambda}$ is some UV cutoff in the theory. The form
immediately suggests the equivalence of massless limit and large magnetic field
limit. This is quite obvious  from the way $B$ and $m$ occur in
$\omega=\frac{eB}{m}$ which is the only energy scale in this problem. If
one compares with $L_0$ then one finds
\begin{equation}
\frac{\Delta f}{L_0}=\frac{2\pi^2 e^2}{3}\ln\left(\frac{eB}{m^2}\right)
\end{equation}
which implies that the radiative correction to the field equations may
become of order of unity in strong magnetic fields given by
\begin{equation}
B_{0}\sim \frac{m^2}{|e|}\exp\left(\frac{3}{8\pi^3\alpha}\right)
\end{equation}
where $\alpha=\frac{e^2}{4\pi}$ is the fine structure constant.
Before we go to discuss the response of the system in an external magnetic
field we would like to make some comments on the response of electric field
alone. Since an external electric field always gives a non-vanishing
probability for pair creation, it is more convenient to discuss the
response to an external magnetic field and calculate the magnetic
permeability $\mu$ rather than $\epsilon$. Then we can deduce $\epsilon$
from $\mu$ if we assume that the vacuum is Lorentz-invariant. If that is
the case then we should have 
\begin{equation}
\epsilon\mu=1
\end{equation}
Depending on the value of $\mu<1$ or $\mu>1$ one can classify the nature
of the vacuum.\\

 From equation (\ref{str}) it is easy to calculate the
magnetisation and the susceptibility in this limit.  The magnetisation is
given by
\begin{equation}
M(B)= -\frac{\pi eB~m^2}{12}-\frac{\pi e^2 B}{6}\ln\left(\frac{eB}{m^2}\right)
\label{mm}
\end{equation}
 Another interesting point before we go to estimate the susceptibility
 is that
there exists a critical magnetic field below which magnetic field will be
totally expelled. This critical field can be estimated as follows. The
effective magnetic field can be defined as $B_{eff}=~B + 4\pi~M$. Here, the
magnetisation $M$ varies according to equation (\ref{mm}).
 Therefore, there exists a
critical magnetic field $B_{c}$ where the effective field $B_{eff}$
vanishes. This critical magnetic field $B_c$ is given by the following equation
\begin{equation}
\ln\left(\frac{eB_c}{m^2}\right)= \frac{6}{\pi e^2}\left(1-\frac{\pi e
m^2}{12}\right) =\tilde{g}(e,m^2)
\end{equation}
and in a more simple form
\begin{equation}
B_c=\frac{m^2}{|e|}\exp(\tilde{g}).
\end{equation}

Therefore,  the susceptibility is 
\begin{equation}
\chi(B)=-\frac{\pi e m^2}{4}-\frac{\pi e^2}{6}~\ln\left(\frac{eB}{m^2}\right)
\end{equation}
Notice that in 3d, the zero field susceptibility has a peculiar
logarithmic form in the free energy.
Even for QED~\cite{jan} 
 one also gets same logarithmic behaviour of the free energy
(and hence in the susceptibility) in this strong magnetic field limit. For
the sake of comparison we notice that
\begin{eqnarray}
\chi_{2d}(B)& \sim & -\frac{1}{\sqrt{B}}\nonumber\\
\chi_{3d}(B)& \sim & -\ln(B)
\end{eqnarray}
This interesting behaviour of the susceptibility in the massless limit is
reminiscent of the fact that the free energy variation in two cases as

\begin{eqnarray}
f_{2d}(B)& \sim & B^{3/2}\nonumber\\
f_{3d}(B)& \sim & B^2~\ln(B)
\end{eqnarray}
Also it is interesting to note that in both cases the free energy is a
non-analytic function of $B$.\\
One can also calculate the critical magnetic field  for which the
effective permeability becomes zero. The permeability is given by
\begin{eqnarray}
\mu(B) &=& 1+4\pi\chi(B)\nonumber\\
~~~~~~~&=&1-\pi^2e m^2-\frac{2\pi^2 e^2}{3}\ln\left(\frac{eB}{m^2}\right)
\label{nb}
\end{eqnarray}
Thus, the critical magnetic field $B_1$ is simply
\begin{equation}
B_1=\frac{m^2}{|e|}\exp(d(e,m^2)),
\end{equation}
where the function $d(e,m^2)$ is determined from the equation
\begin{equation}
d(e,m^2)=\frac{3}{2\pi^2 e^2}\left(1-\pi e m^2\right)
\end{equation}
Thus we have three critical values of magnetic field for which (i)
$B_{eff}$ vanishes, (ii) $\mu(B)$ vanishes and (iii) logarithmic term in
the susceptibility goes away and $\chi(B)$ becomes $-\frac{\pi e m^2}{4}$.
One thing to notice here that all the critical magnetic fields are of 
order $\frac{m^2}{e}$ and these critical fields should be compared with
the field on which the classical theory of electrodynamics will break
down. This classical magnetic field can be estimated from the comparision
of two length scale 
\begin{equation}
\frac{e^2}{m}\sim \frac{m}{eB}
\end{equation}
This gives the critical field
\begin{equation}
B_{class}\sim \frac{m^2}{e^3}
\end{equation}
On comparison we notice that
\begin{equation}
\frac{B_{class}}{B_c}\sim \frac{1}{e^2}
\end{equation}
This is true for all the other critical fields also. Thus, the ratio 
between the classical field and the critical field is of the 
order of fine structure constant.\\

 Next we would like to
compute the dielectric constant $\epsilon(r)$ from the effective permeability.
Since we are working in natural units $\hbar=1$ and $c=1$ then the
dielectric constant $\epsilon$ and the permeability $\mu$ is related by
$\epsilon\mu=1$. However, one can show from the effective action~\cite{eff}
formalism that under certain circumstances one can still use the
relation in presence of an external magnetic field though the physical
situation is not Lorentz-invariant. Hence, the effective permeability
$\mu(B)$ should be related to the dielectric constant $\epsilon(r)$ by the
relation given by
\begin{equation}
\epsilon(r)=\frac{1}{\mu(B)}\mid_{eB\rightarrow 1/r^2}
\end{equation}
This immediately suggests that 
\begin{equation}
\epsilon(r)=\frac{1}{1-\pi^2 e m^2-\frac{4\pi
e^2}{3}\ln\left(\frac{1}{mr}\right)}
\end{equation}
From equation (\ref{nb}) we notice that $\mu(B)<1$ which means that
$\epsilon(r)>1$. This is easily seen from the above equation. Hence, 
the charged scalar field vacuum (diamagnetic) is screening one. Before
we go to finite temperature case  scheme we would like to make a comment
on the vacuum energy of QCD in an external magnetic field. In fact through
this regularisation scheme one can explicitly show the paramagnetic nature
of this spin-1 field~\cite{ja}.

\subsubsection{Finite Temperature Case}
Now, for the finite temperature case  one can regulate the
free energy through the same mode matching regularisation method used
in (2+1) dimension. Finally,
 one can write down the free energy difference in dimensionless form as
before 
\begin{equation} 
\Delta F(B)= F(B)-F(0)=\sum_{l=0}^\infty~d_{l}(b,\delta,\zeta,\rho) 
\end{equation} 
where,
\begin{equation}
d_{l}(\rho,\delta,\zeta)=
\frac{\rho}{\delta}
\int_{-\infty}^{\infty}
dp_z\left[g(\rho,l,1/2)-\int_{0}^{1}~d\alpha~g(\rho,l,\alpha)\right].
\end{equation}
The dimensionless variables are defined as
$\delta=\beta m$ and $\rho=\beta\mu$.

The coefficient $g(\rho,l,\alpha)$ is given by
\begin{eqnarray}
g(\rho,l,\alpha)&=&
\log\left(1-\exp(-\delta(\sqrt{1+2(l+\alpha)\rho}-\zeta))\right)+\nonumber\\
& &\log\left(1-\exp(-\delta(\sqrt{1+2(l+\alpha)\rho}+\zeta))\right)\label{gb1}.
\end{eqnarray}

Now, defining $z_{l}=\frac{(1+2l\rho)}{2\rho}$ we can write the
equation (\ref{gb1}) 
\begin{eqnarray}
 g(\rho,l,\alpha)&=&
\log\left(1-\exp(-(\sqrt{z_{l}+\alpha}-\zeta))\right)+\nonumber\\ &
&\log\left(1-\exp(-(\sqrt{z_{l}+\alpha}+\zeta))\right).
\end{eqnarray}
The function $g(\rho,l,\alpha)$ is convex,  so the zero temperature
argument applies unchanged. It follows that
the free energy satisfies the following inequality
\begin{equation}
F(B) \geq F(0)
 \end{equation} 
 This proves the diamagnetism
for charged scalar fields at finite temperature. It would be interesting
to look into the various limits of this free energy difference. But here
we do not address these limits at all.

\section{Interacting case} In this
section  we want to extend the diamagnetic inequality to the
self-interacting field theory case including the dynamical interaction
between scalar fields. We present here the proof of diamagnetic 
inequality for a generalised d dimensional
case. The partition function of this charged
self-interacting field theory in the presence of the magnetic field can be
written as
\begin{equation} 
{\rm Z(B)}= \int\int{\cal D}[\Phi]{\cal D}[\Phi^\ast]{\rm
exp}(-S(\Phi,\Phi^\ast,A)), \label{fq} 
\end{equation} 
where  the action S is defined as 
\begin{equation}
{\rm S}= \int\int d\,^{d}x~d\tau\left[(D_\mu\Phi)(D^\mu\Phi)^\ast
 +m^2(\Phi^\ast\Phi) +
V(\Phi^\ast\Phi)\right].
\end{equation} 

 The action is not quadratic and $Z(B)$ cannot be evaluated in closed
form. Nevertheless, we can show that the response of the system to an external
magnetic field is diamagnetic.  Since the formal expression for the
partition function may not exist (the integrals may not exist) we impose a
cut off in momentum space. The functional integral in (\ref {fq}) signifies
that one only integrates over those field configurations whose Fourier
transforms have support within a sphere of radius $\Lambda_0$ in 
momentum
space. The partition function then explicitly depends on $\Lambda_0$. 
We do
not explicitly indicate the $\Lambda_0$ and $\mu$ dependence of
$Z(B,\Lambda_0,\mu)$ below.

We divide the action into two parts $S_{0}$ and ${\rm S_{int}}$,
   where $S_{0}$ is the action in the absence of the external field.
    \begin{equation}
  {\rm S}= {\rm S_{0}} + {\rm S_{int}},
  \end{equation}
  where
\begin{eqnarray}
 {\rm S_{0}}&=&\int\int d^{d}x~d\tau 
\left[(\partial_\mu\Phi)(\partial^\mu\Phi)^\ast)+
m^2(\Phi^\ast\Phi)\right.\nonumber\\ & &\left.+ V(\Phi^\ast\Phi)\right],
\end{eqnarray} \\

\begin{eqnarray}
{\rm S_{int}}&=& \int\int d^{d}x~d\tau\left[{-\rm i} e
(\partial_\mu\Phi)(A^\mu\Phi^\ast) +{\rm
i}e(A_\mu\Phi)(\partial^\mu\Phi^\ast)\right.\nonumber\\ 
& &\left.+e^2({\bf A\cdot\bf A})(\Phi\Phi^\ast)\right].
\end{eqnarray}
 
  Notice that $\exp(-S_{0})$ is a
positive measure on the space of field configurations. The ratio
$Z(B)/Z(0)$ can therefore be regarded as the expectation  value of
$\exp(-{\rm S_{int}})$. Since $\exp(-{\rm S_{int}})$ is an oscillatory
function whose modulus is less than or equal to $1$, we conclude that
\begin{equation}
\frac{\rm Z(B)}{\rm Z(0)}  =   \ll {\rm \exp} {-S_{int}} \gg
                                \leq 1
\end{equation}

 This implies that
\begin{equation}
F(B)\geq F(0)
\end{equation}

This result is an {\it exact} and {\it non-perturbative} one. Hence, it is more
general and strong compared to perturbative one.

In this derivation, we have not assumed any form for the vector potential.
So, the result derived above is true for {\it both homogeneous or 
inhomogeneous}
magnetic fields of any strength. Since $\beta$ is arbitrary, the result
holds at {\it all temperatures}. The argument presented here 
works for any
arbitrary interaction $V(\Phi^\ast\Phi)$~( Generally, it is assumed that
$V(\Phi^\ast\Phi)$ is a smooth function, for instance, a polynomial).

\par
Upto now we have considered the cases of charged scalar fields interacting
through a potential. It is also possible to consider interaction mediated
by a dynamical electromagnetic field $A_{\mu}$. The fields in the system
are now $\Phi$ (charged scalar fields) and $A_{\mu}$. If one applies an
external magnetic field $A_{ext}$ then the full Lagrangian is given by
\begin{equation} 
{\cal L}= -\frac{1}{4} F^2 +(D_{\mu}\Phi)^\ast(D^\mu\Phi) -
m^2(\Phi^\ast\Phi) - V(\Phi^\ast,\Phi) 
\end{equation}
 where $D_{\mu} =\partial_{\mu}-{\rm i} e A_{\mu}^{ext}-{\rm i} e A_{\mu}$
and $F_{\mu\nu}=\partial_{\mu}A_{\nu} - \partial_{\mu}A_{\nu}$.

 The
argument given above can be modified as follows. The definition of $S_{0}$
changes slightly while ${\rm S_{int}}$ remains the same.
\begin{eqnarray}
{\rm S_{0}}&=&\int\int d^{d}x~d\tau\left[-\frac{1}{4} F^2+(\partial_\mu-ie
   A_\mu\Phi)^\ast(\partial^\mu+ie A^\mu\Phi)\right.\nonumber\\
   & &~~~~~~\left. + m^2(\Phi^\ast\Phi)+V(\Phi^\ast\Phi)\right],
\end{eqnarray}                  
and
\begin{eqnarray}                
{\rm S_{int}}&=& \int\int d^{d}x~d\tau\left[{-\rm i} e
(\partial_\mu\Phi)(A^\mu\Phi^\ast)\right.\nonumber\\
& &\left. +{\rm
i} e(A_\mu\Phi)(\partial^\mu\Phi^\ast)+e^2({\bf A\cdot\bf
A})(\Phi\Phi^\ast)\right].
\end{eqnarray}
Again one can repeat the same argument to establish the diamagnetic
inequality by noting that $\exp(-S_{0})$ is a positive measure and the
ratio $Z(B)/Z(0)$ as an expectation value of $\exp(-{\rm S_{int}})$. This
universal inequality follows from basic principles and does not depend on
the details of the interaction.

\subsection{Comment on the regularisation scheme in interacting case}
The effective potential $V(\Phi^\ast\Phi)$ is gauge invariant. What we
have done is that we have suitably redefined the measure of the integral
through the inclusion of self-interaction. Thus now, the bare measure has
been changed to so called ``dressed"  measure. This has the advantage that
the effects of self-interaction and mutual interaction have been
separated. Also  In field theory  one would also require that the
interaction $V(\Phi^\ast\Phi)$ be renormalizable. In $(d+1)$
dimensions  this would  restrict the interaction $(\Phi^\ast\Phi)^m$ only.
From simple power counting, one can notice easily that the required value
of $m$ is $\frac{d+1}{d-1}$. In condensed matter physics, where a natural 
cutoff exists, higher order
powers of $(\Phi^\ast\Phi)$ may also be present.\\

Under normal conditions $k_BT\ll m$ we expect the self interaction to
merely affect the particle's mass. This is the same concept one uses in
quasiparticle picture of Landau
Fermi Liquid theory~\cite{qp}. The effect of all the interactions can be
put in the definition of the effective mass of the quasi particles.
Therefore, the error committed would be small if we were to ignore the
self-interactions while keeping each mass at a fixed observalble value.
Hence, we drop the superscript ``dress" and work with the ordinary Wiener
measure.\\

Notice that the bare mass enters the Wiener measure. However, in  the
expression of the ${\rm S_{int}}$ neither refers to nor uses the bare
mass; it never enters the interaction of the scalar particle with the
electromagnetic field. On the otherhand, it is important to realize that
the bare mass $m$ differs from the physical or ``effective" mass
$m_{eff}$. The differnece is induced by the interaction. Adjusting the
bare mass so as to gurantee that $m_{eff}$ agrees with the observed mass
of the particle constitutes the central issue of the renormalization
program. It is generally believed that this issue ought to be resolved
outside the perturbation theory. In this paper we do not attempt to answer
these questions.\\

 In the interacting case it turns out that the partition
function  and hence the free
energy cannot be evaluated in closed form but still one can show the
diamagnetism of charged scalar fields. This shows that diamgnetism of
charged scalar fields is {\it universal}~ i.e. robust in the sense that it
is true for all dimensions  and independent of the interaction between the
scalar fields.  One more comment regarding the meaning of universal
behaviour of this system towards an external magnetic field. The proof of
the diamagnetic inequality in the interacting case does not assume the
form of the vector potential. Thus, in that sense the theory is true
whether the applied magnetic field is homogeneous or inhomogeneous.  The
word {\it universality} also reflects this fact that the behaviour of the
system towards the magnetic field is independent of the nature of an
applied magnetic field. We would like to emphasize one point regarding our
theory which is as follows. If a phase transition takes place changing the
drastic nature of the system, then of course our theory fails. As long as
a phase transition does not take place that alters the bosonic behaviour,
this theory can predict the response of the system in an external magnetic
field as diamagnetic.

\section{Applications}
It is evident from the previous sections that though the results obtained
are quite general and important from the theoretical point of view but
fail to connect directly with the experiment. What we have shown is
basically the nature of the response of a system under an external
magnetic field. Thus, though the results are of interesting one but fails
to compute explicitly the susceptibility as a function of magnetic field,
mass and temperature. Only in zero temperature case we have been able to
compute the susceptibility in two limits. However, one can resort to
perturbation method to compute approximately the response of the system in
case of finite temperature. In this section we would like to discuss the
possible situations which could be well-connected with  the theory
presented. From an experimental point of view one is more interested in
the quantitative variation of the susceptibility with the temperature.
Therefore, in this section we will mainly address the temperature
variation of the diamagnetic susceptibility assuming an effective
Landau-Ginzburg free energy functional apart from giving some illustration
of spinless bose systems. Notice that the order parameter in
Landau-Ginzburg model is a scalar one which has a characteristic $U(1)$
symmetry. Only the mass parameter in Landau-Ginzburg model has a different
meaning in contrast to the theory presented here.\\

 One obvious example spinless bose systems in the laboratory is the Cooper
pair formation in superconductors which shows  perfect diamagnetism (known
as the Meissner~\cite{sup} effect) below the critical temperature. Cooper
pairs also exist in Neutron Stars~\cite{neutr} where the magnetic field is
very high compared to any laboratory  field. Of course, the operators
which create and destroy Cooper pairs are not strictly Bose operators, so
this is only an analogy. In fact one can also calculate the diamagnetic
susceptibility of the Cooper pairs above the critical temperature, more
specifically the temperature variation of the susceptibility. The diamagnetic 
susceptibility above the critical temperature has been observed recently in
High-$T_c$ compounds~\cite{exp}. In literature there exists a
model~\cite{bo} which does predict that above the critical temperature all
or at any rate some of the carriers are bosons. Here, of course, as has
been pointed out~\cite{mott} that the definition of the ``critical
temperature" bears a different meaning in contrast to usual one. The
``critical temperature" in that context refers to the temperature at which
the gas of bosons become wholely {\it non-degenerate}. However, in the
following we do not address all these issues regarding the mechanism of
formation of bosons.\\

 With the usual Landau-Ginzburg effective field
theory~\cite{LG} it turns out that the diamagnetic susceptibility 
in $d$ dimension
$\chi_d\sim -(T-T_c)^{-\frac{4-d}{2}}$.  Without going  much details into
this calculation~\cite{dbj}, the above peculiar variation of the 
susceptibility with
temperature can be explained as follows. Following the argument given for
3d~\cite{ma}, we notice that above $T_c$, droplet of Cooper pairs will
grow and decay as a result of thermodynamic fluctuations. Their mean
radius is approximately equal to $\xi$ which is the coherence length
of the Cooper pairs and phenomenologically the simplest variation
of this coherence length is taken as $\xi^2\sim
(T-T_c)^{-1}$. The ``mass parameter" in this theory is inversely proportional
to this coherence length. Therefore, the amount of energy required to produce a
droplet is given by
\begin{equation}
\delta E\sim \frac{1}{\xi}\times \xi^d\times |\phi|^2
\end{equation}
Here $|\phi|^2$ is the density  of the Cooper pairs in the framework of
phenomenological Landau-Ginzburg free energy functional.  This energy must
be equal to the thermal energy $k_BT_c$ and hence
\begin{equation}
|\phi|^2\sim \frac{k_BT_c}{\xi^{d-1}}
\end{equation}
Now consider the expression for the diamagnetic susceptibility of atoms in
scaled form as
\begin{equation}
\chi_{d}\sim -\frac{|\phi|^2 e^2\xi^2}{1/\xi}\sim
-(T-T_c)^{-\frac{4-d}{2}}
\end{equation}
In otherwords for a general variation of the coherence length
$\xi^2\sim (T-T_c)^{-\nu}$ the susceptibility variation turns out
as $\chi_d\sim -(T-T_c)^{-\nu(4-d)/2}$. 
However, as it is evident for dimension  $d\geq 4$ the above formula does
not make any sense; but this formula of course correctly reproduces the
expected temperature variations for 2d and 3d. In fact, in 3d the exponent
of the susceptibility ( $1/2$ ) has been observed in the
experiment~\cite{exp}. Notice that inspite of the strong correlation among
the Cooper pairs in the system the pairs behave as if they are
non-interacting. Thus, within  this simple Landau-Ginzburg approach one can
give a strong hint towards the role of dimensionality of the system in
High-$T_c$ material.\\

It has been shown explicitly by J. Daicic et.
al.~\cite{mag} that the magnetised pair Bose systems are relativistic
superconductors. These systems are not covered by previous analysis
~\cite{van,lan,sim,ss} which apply to non-relativistic quantum mechanical
systems. Pions would be  suitable candidates for application of this
theory with $\pi^+$ and $\pi^-$  regarded as the particles and
antiparticles. The choice is also motivated by the fact that these pions
are massive ($mc^2=139.5673$ Mev), obey Bose-Einstein statistics and that
they possess no spin. It is also well known that hard core
bosons~\cite{bss} in any dimension on any lattice show a preference for
zero flux.

\section{Conclusions and Perspectives}
\par
 The response of a system to an electric field is completely different
 from its response to a magnetic field.       
 The basic difference between the responses of a system on  
 application of an electric field
 or a magnetic field lies in the Hamiltonian of the system.      
 
 The Lagrangian of a system in the presence of an electric field
 can be written as
 \begin{equation}
 {\cal L}=(D_{0}\Phi)^\ast(D_{0}\Phi) -
(\nabla\Phi)^\ast(\nabla\Phi)-m^2(\Phi^\ast\Phi)- V(\Phi^\ast\Phi)
\end{equation}
 where 
\begin{equation}
  D_{0} = \partial_{0}-{\rm i} e A_{0}
\end{equation}
 For statistical mechanics to make sense, the Hamiltonian
$H$ must be independent of time. Then it follows that
 \begin{eqnarray}
{\cal H}&=&(\Pi^\ast)(\Pi)+(\nabla\Phi)^\ast(\nabla\Phi)+
m^2(\Phi^\ast\Phi)+V(\Phi^\ast\Phi)\nonumber\\
& &-{\rm i}e\left[(\Pi^\ast)(A_{0}\Phi)-(\Pi)(A_{0}\Phi^\ast)\right]
\end{eqnarray}\\
 
The electric field appears in the Hamiltonian through the linear vector
potential $A_{0}$ term. Now, from finite temperature second order
perturbation~\cite{lan1,lan2} theory, one can show  easily that the free
energy of the system always decreases with the electric field. Hence, the
dielectric susceptibility is always positive in thermal equilibrium.\\

But in the case of a magnetic field the Hamiltonian contains
both linear and quadratic terms in $A$. The net effect of an applied
magnetic field is not {\it a priori} clear. However, as our analysis makes
clear, for charged scalar field theories the net effect is always
diamagnetic.\\

In case of spinless bosons, there is no Zeeman term coupled
with a magnetic field and hence the system consisting of spinless bosons
always has higher energy in a magnetic field than without the magnetic
field. It has been already pointed~\cite{sim,ss,bss,dav} out that there is
no corresponding theorem for fermions. So, in case of fermions having spin
the general tendency is to show paramagnetism~\cite{para}.\\
  
 Before we end we would like to comment on a recent work in the literature
\cite{Kraemmer} on  ultrarelativistic hot scalar plasma. The results of
their perturbative treatment are consistent with our results. In contrast,
as we have already pointed out that our treatment is a non-perturbative
one.  Since any non-perturbative treatment is always welcome in field
theory we thought it would be interesting to present these results and
arguments.  Besides we have been able to show the renormalisation of the
free energy.  These regularisation schemes help one to understand various
interesting renormalization of the quantity involved (in this case charge)
and a proper justification of the scaling functions discussed in the
paper.\\

 In  summary, we have shown exactly, that  charged
scalar fields in (2+1) and (3+1) dimension at all  
temperatures are diamagnetic.\\

\noindent
\acknowledgements

It is a pleasure  to thank  Joseph  Samuel for suggesting this problem
and stimulating discussions; Narendra Kumar, Rajaram Nityananda, Supurna Sinha,
G. Kang, Chandrakant S Shukre, Diptiman Sen and Rajesh N. Parwani
for useful discussions.


\newpage
\appendix
\section{Comment on an another regularisation scheme}
In this appendix we would like to discuss about another regularisation
scheme which at first sight one might think of as a correct 
regularisation
scheme. However as we show below this method is not a correct scheme.
This scheme can be thought of
 as relating $L$ and $\Lambda$ by the
criterion that the maximum energy of the two systems should be matched.
 The matching of the energy cutoff gives the relation between the
momentum cut off $\Lambda$ and the Landau level $L$ as
\begin{equation}
E_{L}=\sqrt{m^2+(2L+1)eB}=\sqrt{\Lambda^2+m^2}
\end{equation}
This immediately suggests that
\begin{equation}
\Lambda^2=(2L+1)eB
\end{equation}
Note that this relation is different from the mode matching
relation(\ref{mode1}) we used previously for the charged scalar field theory.
However, it turns out that this renormalisation procedure gives a divergent
answer for the free energy difference $\Delta f$. We support this
statement through both numerical as well as analytical calculations.\\

Below we present numerically a comparison between the two 
schemes.  The values of $\Delta f^A$ (scheme due to anonymous) and $\Delta
f^J$ (used in the paper) are tabulated for $B=2$, $m=.1$ and for various
values of the cutoff $L$.\\

\begin{tabular}{ccc}
\tableline
\tableline
Different values of cutoffs($L$) &  ~~$\Delta f^A$ &  ~~$\Delta f^J$\\
\tableline
L=10&~~6.774&~~0.2157\\
L=100&~~20.308&~~0.2325\\
L=1000&~~63.507&~~0.2382\\
L=10000&~~200.248&~~0.2400\\
L=50000&~~447.457&~~0.2404\\
L=80000&~~565.929&~~0.2405\\
L=100000&~~632.698&~~0.241\\
\tableline
\tableline
\end{tabular}\\

From this table it is easy to see that our  method gives a
finite convergent, cutoff independent answers while the other method gives
cutoff dependent, divergent answers for the free energy. In fact one can
show analytically in this case that $\Delta f$ diverges with the cutoff as
$\sqrt{L}$. This unphysical and meaningless answer is due to the
comparision of systems with different number of degrees of freedom. Note
that in our analysis $\Delta f$ converges as $\frac{1}{\sqrt{L}}$.  Hence,
the scheme due to matching of highest energy is therefore clearly
unphysical. However, as already discussed our scheme can be justified
physically from the adiabtically turning on the magnetic field. Then the
energy of each mode is affected by the magnetic field.  Comparing systems
with the same number of modes is the logically and physically correct
procedure.  What the other method does is that it compares the systems
with different degrees of freedom. Hence, regularisation scheme based on
energy cutoff does not give a renormalisable answer. Also notice that
phase space is invariant in any Lorentz frame while the energy is not. We
present below  analytical work to show that in our case $\delta f$ varies
with $\frac{1}{\sqrt{L}}$ and as $L\rightarrow\infty$, it does converge
while the other case it diverges as $\sqrt{L}$ in the same limit as
indicated above.  Since the ultimate aim is to take the cutoff to infinity
let us calculate the behaviour of the leading term in free enrgy
difference as a function of cutoff.  The difference between the free
energy (using the two regularisation schemes) is given by
\begin{eqnarray}
\delta f_{L}(B,m)&=&\frac{1}{3}\left[(m^2+(2L+1)eB)^{3/2}\right.\nonumber\\
& &\left.-(m^2+2(L+1)eB)^{3/2}\right]
\end{eqnarray}
Now defining a dimensionless small parameter $\eta=\frac{m^2}{(2L+1)eB}$
(which is reasonable in large cutoff $L$ and in strong magnetic field) it
is easy to write the above difference as
\begin{equation}
\delta f_{L}(\eta) \sim \sqrt{L}+ O(\frac{1}{\sqrt{L}})
\end{equation}
This is consistent with the numerical results presented in the
above table.
Another way of qualitatively understanding the above divergence is the
following . In a strong magnetic field limit all the higher
Landau levels are far away and one is interested only in the lowest Landau
level. Now in this situation if one countes the energy states and matches
with the free case, then there will be leading order diveregnce with the
cutoff due to mismatch of  number of degrees of freedom. Thus we
notice that though the regularisation scheme provides a finite positive
free energy differenece, it fails to give a renormalised difference.
Hence this scheme should not be considered as a correct and physical
regularisation prescription.
\section{Uniqueness of the regularisation scheme used}
In this appendix we  want to  show the uniqueness of the
regularisation method used in this paper.
In particular, we would like to argue that the mode matching relation
(\ref{mode1}) can be obtained from the condition that the difference
between the two free energies is independent of the cutoff and tends to
some finite value. Here too we assume that the difference is finite for
all values of $B>0$ and $m>0$. As we have noted before for charged scalar
fields the difference between the free energies can be written as
\begin{equation}
\Delta f= eB\sum_{l=0}^L\sqrt{m^2+(2l+1)eB}-\int_{0}^{\Lambda}
~pdp~\sqrt{p^2+m^2}\label{ws}
\end{equation}
We want to find the relation between $\Lambda$ and $L$ for which the
difference tends to a finite value independent of the cutoffs. In
otherwords as $L\rightarrow\infty$ or $\Lambda\rightarrow\infty$ the
difference between the free energy becomes unique and finite. Though in
principle $\Delta f$ depends on both $\Lambda$ and $L$ separately, we
will notice that because of a unique relation between them we will get a
finite value independent of both of them. Hence the free energy difference
can be written as
\begin{equation}
\Delta f=\Delta f_0+ O(\frac{1}{g(L)}),
\end{equation}
where the first term is the universal component which is independent of
the cutoff (its value of course depends on $eB$ and $m$) and the second
term is the correction to it. The function $g(L)$ is assumed to be a
smooth function of $L$ and in the present analysis we are not interested
in the actual form of it. We assume the cutoff to be high enough so that
we can replace the sum by an integral. Now if we demand that $\Delta f_0$
should be independent of the cutoff then we get an equation of the form
\begin{equation}
\Lambda\frac{\partial \Lambda}{\partial L}\sqrt{m^2+\Lambda^2}\rightarrow
eB \sqrt{m^2+2L eB}
\end{equation}
We have used the symbol $\rightarrow$ instead of an equal sign to signify
``asymptotically approaching" rather than equality between the two
relations. In fact they are strictly equal only if the cutoff is taken to
infinity. But, because of finite value of  $L$ or $\Lambda$
however large, we do not expect strict equality in the above relation.
From this equation in the first order approximation it is quite evident
that there should be a relation between $\Lambda$ and $L$ such that
\begin{equation}
\Lambda\frac{\partial \Lambda}{\partial L}= eB
\end{equation}
This immediately gives a relation between the two cutoff within an
undetermined constant $C$ as
\begin{equation}
\Lambda^2=2eBL+2C
\end{equation}
It is important to notice that this relation has been obtained in the
limit of large enough cutoff. In fact as we have seen in the previous
subsection the relation is valid for small as well as large value $L$.
However, the physical energy difference approaches to finite value as 
 $\frac{1}{\sqrt{L}}$ as  we vary $L$.
In the next we will try to fix this constant from the fact that the
difference goes to a finite value. We substitute this value in the free
energy difference equation (\ref{ws}) and plot the difference as a
function of various values of $L$ for fixed values of B and m. 
 Then it turns out 
 that the only allowed value of $C$ for the free energy to be
finite and independent of the cutoff is $eB$. Thus we find that the exact
relation between the two cutoff is
\begin{equation}
\Lambda^2=2eBL+2eB=2eB(L+1)
\end{equation}
It is interesting to note that no other values of $C$ are allowed for the
free energy difference to be finite and independent of cutoff.
Thus we have justified the mode matching relation (\ref{mode1}). In fact,
this is the only method by which one can get a unique finite
cutoff-independent free energy difference.


\begin{references}
\bibitem{scale} The explicit form of this scaling function can be obtained if
one rugularises through mode matching scheme.
\end{references}
\end{document}